\documentclass[%
twocolumn,
superscriptaddress,
 amsmath,amssymb,
 aps,
prl,
]{revtex4-2}

\usepackage{graphicx}
\usepackage{dcolumn}
\usepackage{bm}
\usepackage{color}

\begin{document}

\preprint{APS/123-QED}
\title{Compact Slow-Wave Hybrid Magnonics on Ultrahigh-Permittivity Substrates}

\author{Yu Jiang}
\thanks{These authors contributed equally to this work.}
\affiliation{ 
    Department of Electrical and Computer Engineering, Northeastern University, Boston, MA 02115, USA
}

\author{Haiyang Yun}
\thanks{These authors contributed equally to this work.}
\affiliation{ 
    Department of Electrical and Computer Engineering, Northeastern University, Boston, MA 02115, USA
}

\author{Zixin Yan}
\affiliation{ 
    Department of Electrical and Computer Engineering, Northeastern University, Boston, MA 02115, USA
}

\author{Amin Pishehvar}
\affiliation{ 
    Department of Electrical and Computer Engineering, Northeastern University, Boston, MA 02115, USA
}

\author{Benyamin Davaji}
\email{b.davaji@northeastern.edu}
\affiliation{ 
    Department of Electrical and Computer Engineering, Northeastern University, Boston, MA 02115, USA
}
\affiliation{ 
    Department of Physics, Northeastern University, Boston, MA 02115, USA
}

\author{Xufeng Zhang}
\email{xu.zhang@northeastern.edu}
\affiliation{ 
    Department of Electrical and Computer Engineering, Northeastern University, Boston, MA 02115, USA
}
\affiliation{ 
    Department of Physics, Northeastern University, Boston, MA 02115, USA
}

\date{\today}


\begin{abstract}

Hybrid magnonics based on spoof surface plasmon polaritons (SSPPs) enables strong, broadband magnon-photon coupling, but the centimeter-scale device footprints required at low gigahertz frequencies hinder integration. Here, we demonstrate miniaturized slow-wave hybrid magnonic devices using an ultrahigh-permittivity ceramic substrate ($\varepsilon = 140$). Owing to the inverse-square-root dependence of the SSPP plasma frequency on effective permittivity, the required corrugation finger length is reduced by nearly a factor of four compared with conventional microwave substrates. The ceramic is thick and brittle, and conventional thin-film processes cannot easily pattern thick, low-loss conductors on it. We overcome this challenge with additively manufactured electronics (AME). In a customized AME workflow, a multi-material inkjet printer first prints a dielectric fixture that holds and registers the ceramic. After the ceramic is loaded, the printer prints thick silver SSPP patterns directly onto it. Measurements on six printed devices with different finger lengths confirm the predicted geometry-frequency scaling. Moreover, the enhanced field confinement of the compact SSPP waveguide nearly doubles the magnon-photon coupling strength and yields ultrastrong coupling with $g/f_\mathrm{p} \approx 22\%$. Notably, this performance is achieved using a YIG film whose magnetic volume is more than an order of magnitude smaller than that employed in previous SSPP demonstrations. These results establish AME on ultrahigh-permittivity substrates as a practical route toward compact, strongly coupled hybrid magnonic devices.

\end{abstract}

\maketitle


Hybrid magnonics has emerged as a versatile platform for coherent information
processing, in which magnons---the quantization of collective spin excitations in magnetically
ordered materials---interact coherently with photons, phonons, and other
excitations~\cite{Rameshti_PhysRep_2022, Harder_SSC_2018, YiLi_JAP_2020, Awschalom2021Feb, Lachance_APE_2019, Bhoi_SSP_2020, Yuan_PR_2022Jun, Huebl_PRL_2013, XufengZhang_PRL_2014}. Owing to the low damping of yttrium iron
garnet (YIG) and the wide tunability of the magnon frequency, these systems
allow coherent information exchange among different degrees of freedom, enabling functionalities ranging
from microwave-to-optical transduction~\cite{Hisatomi_PRB_2016, NaZhu_Optica_2020, Osada_PRL_2016, XufengZhang_PRL_2016}
and quantum sensing~\cite{Tabuchi_science_2015Jul, Lachance-Quirion_Science_2020Jan, Xu_PRL_2023May} to nonreciprocal
signal control~\cite{YipuWang_PRL_2019, XufengZhang_PRAppl_2020} and dark-matter
detection~\cite{Crescini_PRL_2020, Flower_PhysDarkUniv_2019}. Translating these capabilities into
practical technologies increasingly calls for compact, chip-compatible
magnonic circuits, motivating a broad effort to miniaturize magnonic devices
and their microwave interfaces~\cite{Chumak_IEEETransMagn_2022, Flebus_roadmap_2024, wang2020magnonic, Wang_PRAppl_2024, cocconcelli2024tuning, Zhang2023Sep_MTE}.

Among various hybrid magnonic platforms, slow-wave hybrid magnonics holds a unique position because it simultaneously allows broadband operation and enhanced interaction, circumventing the drawbacks of previous configurations based on resonators or conventional waveguides. By employing spoof surface plasmon polaritons (SSPPs)---engineered
surface waves on periodically corrugated microstrips, the microwave signals are significantly slowed down near a geometry-defined cutoff~\cite{Pendry_Science_2004, Gao_AdvMater_2018, Garcia-Vidal_RMP_2022, Tang_AOM_2019, Shen_PNAS_2013, Jensen_OptCommun_2008Nov, Zang_PRAppl_2016Feb}. The reduced group velocity, and the consequent
tightened field confinement, dramatically enhance the magnon-photon interaction
while preserving a broad operating bandwidth~\cite{2024_PRL_Xu_slowwave}, enabling new opportunities such as broadband magnomechanics
in YIG~\cite{jiang2025broadband, 2021_PRAppl_Xu_pulseEcho}.

However, in existing demonstrations, the lateral dimensions of SSPP waveguides are set by the operating wavelength and can reach the centimeter scale at low-gigahertz frequencies, limiting further device integration. High-permittivity substrates offer a route to miniaturization by reducing the electromagnetic wavelength and hence the lateral dimensions of the microwave structure~\cite{Huang_MOTL_2007, Sebastian_JEuCeramSoc_ltcc}. The stronger dielectric loading also tightens the field confinement, providing an additional means of enhancing the magnon–SSPP interaction. Here, we replace the conventional microwave substrate Rogers TMM10i ($\varepsilon=9.8$) with an ultrahigh-permittivity ceramic ($\varepsilon=140$), reducing the SSPP finger length by approximately a factor of four while further enhancing the magnon–SSPP coupling. To fabricate devices on this substrate, we adopt additively manufactured electronics (AME), a maskless process that prints conductive and dielectric inks layer by layer and has recently been used to produce high-quality microwave passive components~\cite{McKerricher_MSNE_2017, Tan_AM_review} and multi-material microsystems~\cite{Arzefouni_PowerMEMS_2025, Yun_PowerMEMS_2025}. AME enables direct metal–ceramic contact to ensure strong dielectric loading of the SSPP mode, together with conductor thicknesses exceeding the microwave skin depth to suppress ohmic loss in the slow-wave regime. These requirements are difficult to satisfy simultaneously using conventional thin-film deposition or transfer-based approaches.


\begin{figure}[tb]
\includegraphics[width=\linewidth]{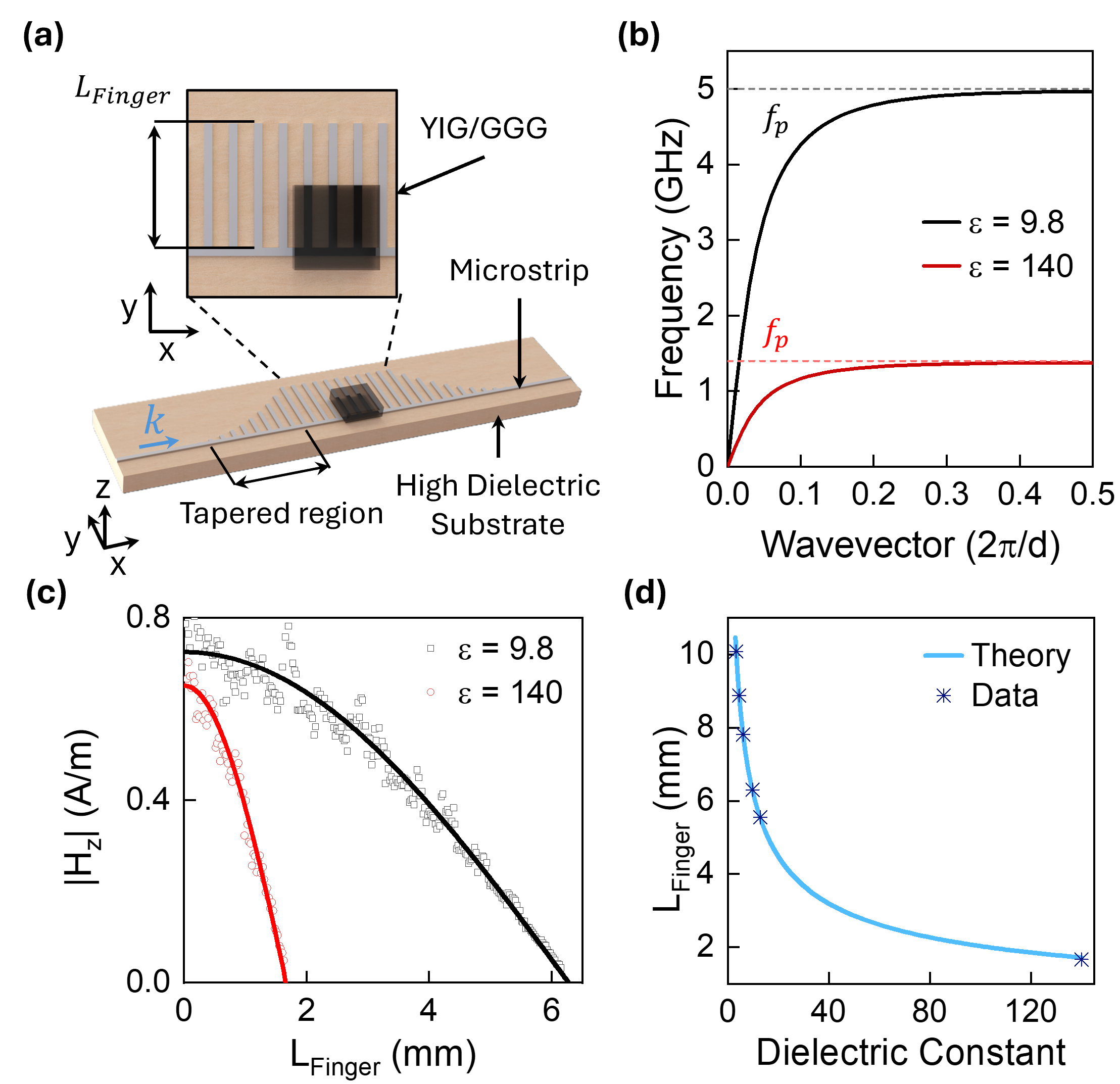}
\caption{ Concept of substrate-enabled miniaturization for slow-wave hybrid magnonics. (a) Schematic of the SSPP-magnonic device. The inset shows the corrugated SSPP region loaded with a YIG/GGG chip. (b) Simulated dispersion relations of the fundamental SSPP mode on substrates with $\varepsilon=9.8$ and $140$. The finger length is 6.3 mm in both cases. Dashed lines indicate the corresponding plasma frequencies. (c) Simulated $H_z$ field profiles along the $y$ direction for $\varepsilon=9.8$ and $140$. Solid curves are cosine fits. (d) Finger length required to achieve a fixed plasma frequency as a function of substrate permittivity. The solid curve is obtained from COMSOL simulations, and symbols denote experimentally measured values for the Rogers and K140 devices. }
\label{Fig1}
\end{figure}

Our device is shown schematically in Fig.~\ref{Fig1}(a). The SSPP waveguide is an array of metallic finger structures (length $L$) with one end attached (shorted) to a microstrip waveguide while the other end open. It supports propagating slow-wave microwave signals that mimic optical surface plasmons. The SSPP waveguide is connected to a regular microstrip on each end through a transition region to suppress reflection. The magnonic resonator is an YIG film (thickness: 23~$\mu$m, size: $3\times3$~mm$^2$) grown on a gadolinium gallium garnet (GGG) substrate, biased by a magnetic field along $y$ direction. The YIG chip is flipped onto the SSPP waveguide near the shorted end of the metallic fingers, where the magnetic field of the SSPP mode is maximum. Thanks to the reduced group velocity and the traveling wave nature of the SSPPs, enhanced magnon-photon interaction can be obtained within a broad bandwidth.

The SSPP waveguide is characterized by its effective plasma frequency $f_\mathrm{p}$. As shown in Fig.~\ref{Fig1}(b), when the mode frequency approaches $f_\mathrm{p}$, the slope of the dispersion curve gradually decreases, corresponding to reduced group velocity and accordingly enhanced magnon-SSPP coupling. In the quasi-static limit, $f_\mathrm{p}$ is governed jointly by the corrugation geometry and the permittivity of the supporting substrate \cite{2024_PRL_Xu_slowwave}
\begin{equation}
f_\mathrm{p} = \frac{C}{4\, L \sqrt{\varepsilon_\mathrm{eff}}} ,
\label{eq:fp}
\end{equation}
where $C$ is a geometric constant and $\varepsilon_\mathrm{eff}$ is the effective permittivity experienced by the mode. For a given geometry, the plasma frequency obtained on a substrate with higher relative permittivity is lower than that on a low-permittivity substrate, which is consistent with the numerical simulation results shown in Fig.~\ref{Fig1}(b). Consequently, to achieve the same plasma frequency $f_\mathrm{p}$, a much shorter finger length is required when a high-permittivity substrate is employed. For example, for $f_\mathrm{p} = 5$~GHz, the required finger length is only 1.66~mm for the substrate with $\varepsilon = 140$, whereas a finger length of 6.30~mm is needed for the Rogers substrate with $\varepsilon = 9.8$. Therefore, the use of the high-permittivity substrate enables nearly a fourfold reduction in the lateral dimension of the SSPP waveguide. This miniaturization also leads to substantially enhanced lateral confinement of the SSPP mode, as confirmed by the simulated mode profiles shown in Fig.~\ref{Fig1}(c). 

The scaling behavior predicted by Eq.~(\ref{eq:fp}) is quantified by the solid curve in Fig.~\ref{Fig1}(d), which shows the finger length required to maintain a fixed plasma frequency $f_\mathrm{p}$ as a function of substrate relative permittivity, as obtained from COMSOL simulations. The study encompasses commercially available substrates spanning a broad permittivity range, from conventional microwave laminates to high-permittivity ceramics with $\varepsilon = 140$. The results reveal a strong inverse dependence of the required finger length on substrate permittivity, demonstrating that high-permittivity substrates provide a practical and effective route toward the miniaturization of SSPP devices.


Realizing such devices, however, requires patterning conductive metallic structures directly on our high-permittivity substrate (K140), which is a thick and brittle ceramic with $\varepsilon=140$. To fully leverage the advantages of SSPPs, several fabrication constraints must be satisfied. First, the conductor thickness must substantially exceed the microwave skin depth, which is on the order of a micrometer, to suppress ohmic loss. As a result, conventional thin-film deposition techniques, such as sputtering and evaporation, are generally unsuitable. Second, the conductor must remain in intimate contact with the substrate to maximize the enhancement provided by its high permittivity. Any intervening gap or thick adhesion layer can reduce this interaction, limiting the effectiveness of transfer-based or flip-chip integration approaches. To address these challenges, we employ AME based on three-dimensional (3D) circuit printing. This maskless additive process deposits both the supporting structure and the conductor in a unified workflow. This approach enables direct fabrication on ceramic substrates and rapid exploration of SSPP designs with varying finger lengths, providing a practical route toward compact SSPP devices based on high-permittivity materials.

\begin{figure}[tb]
\includegraphics[width=\linewidth]{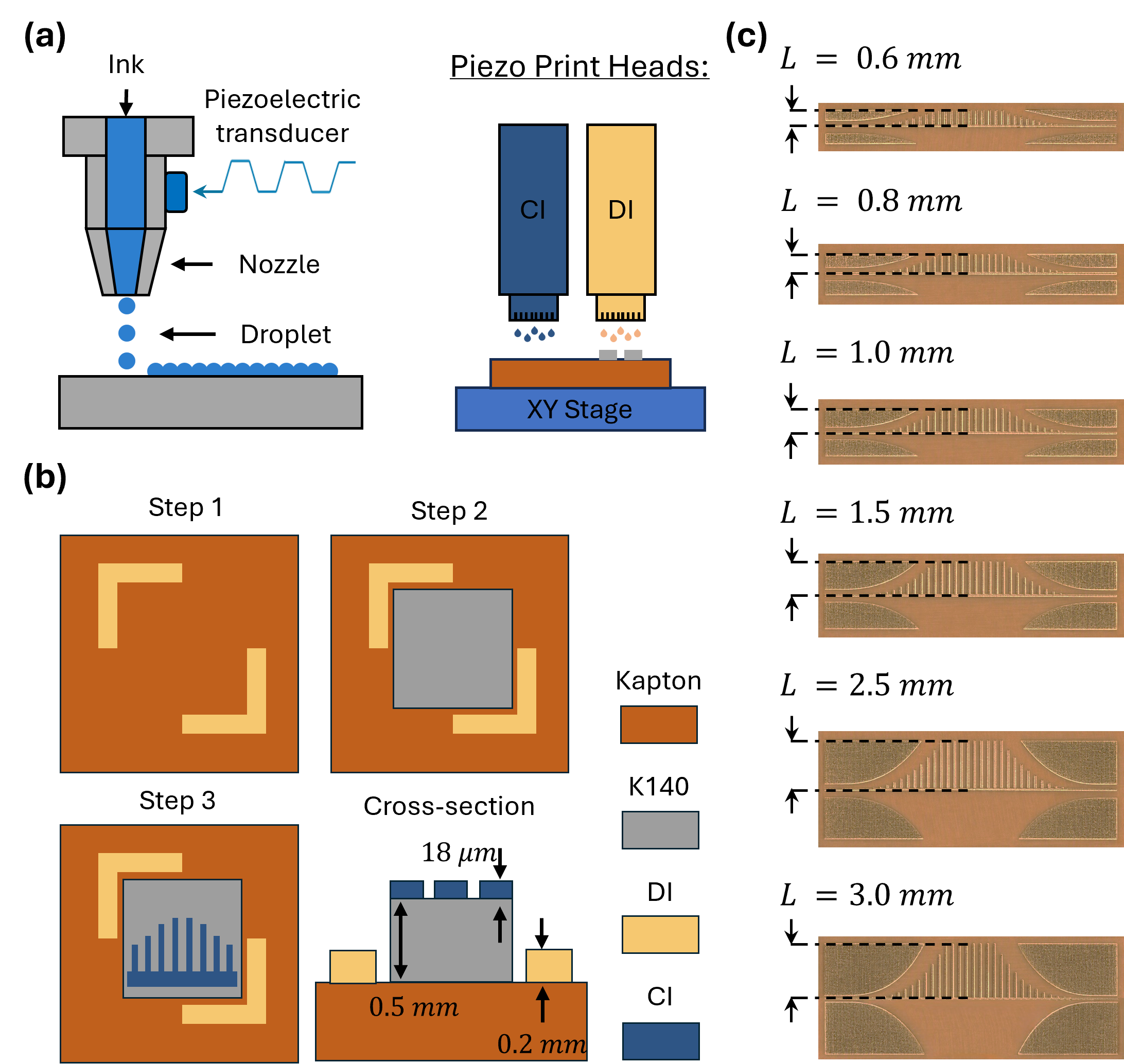}
\caption{AME of SSPP devices by piezoelectric inkjet printing. (a) Schematic of drop-on-demand piezoelectric inkjet printing (left), with two print heads dispensing the conductive ink (CI) and dielectric ink (DI) onto a substrate on the XY stage (right). (b) Customized AME workflow on the K140 substrate. Step 1: two L-shaped DI corner frames are printed on the Kapton sheet. Kapton is the default flexible print substrate and is held by the vacuum stage. Step 2: the job is paused, and the 1 in $\times$ 1 in K140 substrate is placed between the frames. Step 3: the SSPP pattern is printed in CI directly onto the K140. The DI frames hold the substrate only laterally, so the finished device is released by lifting it out of the fixture after Step 3. Cross-section: schematic after Step 3 (not to scale), showing the 0.5-mm-thick K140, the 0.2-mm-high DI frames, and the 18-$\mu$m-thick CI pattern. (c) Optical micrographs of the six printed devices, which differ only in maximum finger length: $L_1$--$L_6=600$, 800, 1000, 1500, 2500, and 3000 $\mu$m.}
\label{Fig2}
\end{figure}

The SSPP devices are fabricated using a DragonFly IV AME system that operates by drop-on-demand jetting [Fig.~\ref{Fig2}(a)] ~\cite{NanoDimension_DragonFly_2025, Arzefouni_PowerMEMS_2025, Yun_PowerMEMS_2025}. A voltage waveform applied to a piezoelectric transducer deforms the ink chamber and ejects a single droplet per pulse, so that both the droplet volume and the jetting frequency are set electronically rather than by nozzle geometry alone. The system carries two independent print heads: one dispenses a silver-nanoparticle conductive ink (CI), and the other a polymer dielectric ink (DI), each with its own reservoir pressure, jetting waveform, and print speed. The platform resolves a minimum trace width and edge-to-edge spacing of 100~$\mu$m, providing sufficient spatial resolution for our SSPP waveguide design where the smallest features are at least several hundreds of micrometers.

A practical obstacle is that the printer is a closed system: by default, material is deposited onto a flexible Kapton sheet held by the vacuum stage of the printer, and the machine provides no native mode for printing onto an externally supplied substrate. This poses a practical limitation for our device fabrication, which specifically requires the high-permittivity K140 ceramic. To resolve this, we implement a customized AME workflow built on a manufacturer-supplied control extension that exposes the layer-by-layer print sequence and permits a programmed pause at an arbitrary layer index. During the pause the build volume becomes accessible, so the ceramic substrate can be loaded without aborting the job or losing the printer coordinate frame, thereby preserving registration of the pattern to the substrate across the interruption.

Our workflow proceeds in three steps [Fig.~\ref{Fig2}(b)], with the DI fixture and the CI pattern cured under separate thermal profiles appropriate to each ink chemistry. First, two L-shaped corner frames are printed in DI on the Kapton sheet over 20 layers at a nominal layer thickness of 10~$\mu$m, giving a frame height of approximately 200~$\mu$m, and cured with the build bed held at 120~$^\circ$C and a UV lamp cross-linking each layer. The two frames sit at opposite corners and define a square pocket, printed in the same coordinate frame that will later position the K140 substrate, which ensures precise alignment of different structures in the SSPP pattern. Second, the job is paused and the K140 substrate ($0.5\times 25\times25~\mathrm{mm}^3$ in size) is placed at the center of the pocket, where the inner walls of the corners provide lateral constraint to fix the substrate position steadily. Third, the SSPP pattern is printed in CI directly onto the ceramic over 18 layers at 1~$\mu$m per layer, giving a final conductor thickness near 18~$\mu$m, with the build bed raised to 160~$^\circ$C and each pass cured immediately by an infrared source to suppress void formation before a final extended exposure sinters the film into a continuous conductor. The nozzle standoff is raised to 800~$\mu$m for this step to clear the elevated substrate surface. After Step 3, the K140 substrate is lifted directly out of the pocket to release the device. This is possible because the DI frames constrain the substrate only laterally.

Following this procedure, six devices were printed and inspected on a digital optical microscope to verify the trace width, gap width, and finger length against the design values [Fig.~\ref{Fig2}(c)]. The devices share an identical unit cell (period: 300~$\mu$m, finger width: 100~$\mu$m) and differ only in their maximum finger length, which ranges from 600 to 3000~$\mu$m, corresponding to different cut-off frequencies. This set of controlled variations provide a direct validation of the miniaturization relation of Eq.~(\ref{eq:fp}): by sweeping the finger length alone, the devices provide a direct experimental map between geometry and cutoff frequency, which we characterize next.

\begin{figure}[tb]
\includegraphics[width=\linewidth]{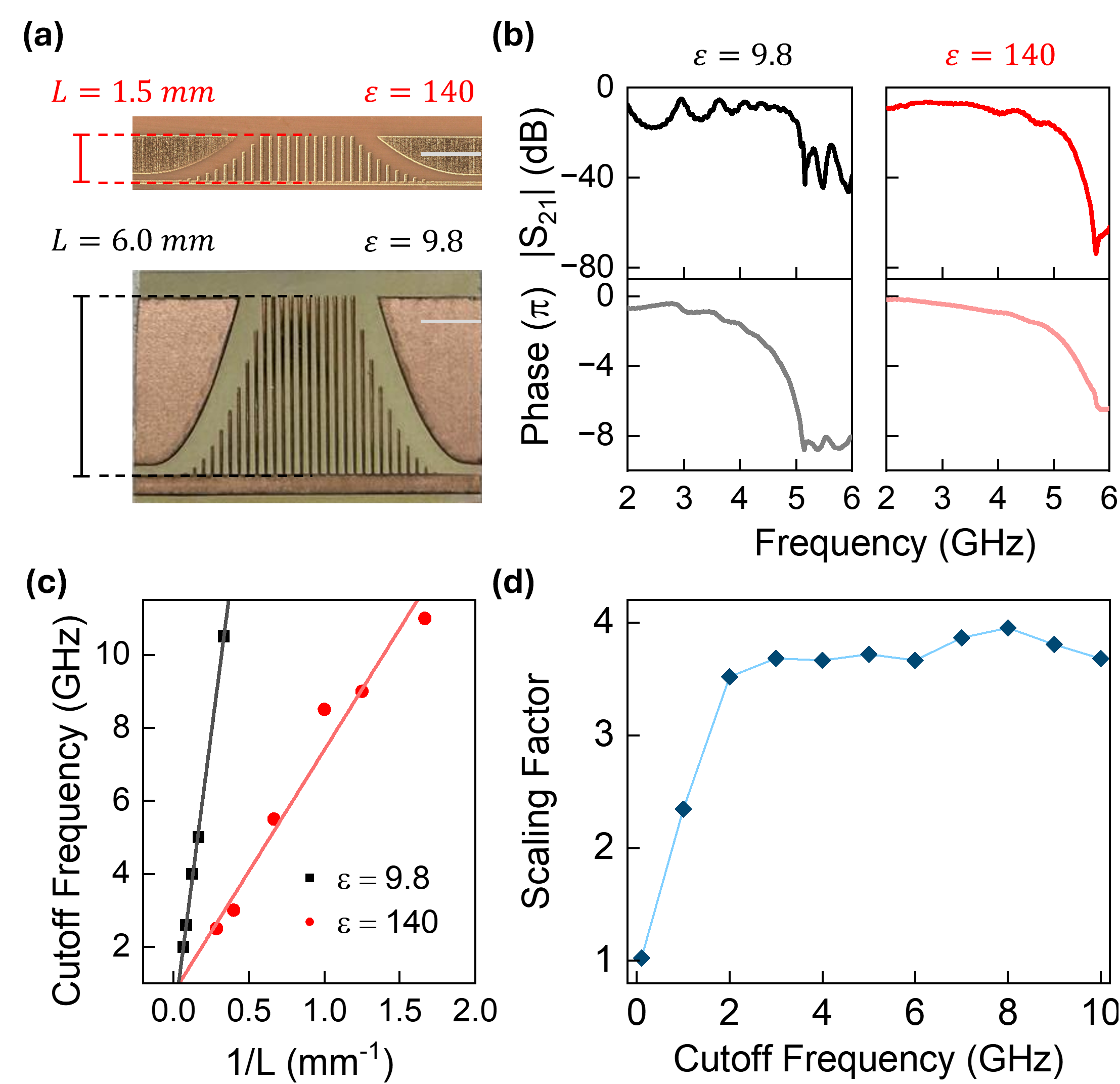}
\caption{ Experimental characterization of SSPP devices on Rogers and K140 substrates. (a) Optical images of SSPP devices designed for the same plasma frequency $f_\mathrm{p}\approx5$~GHz. Scale bars: 2~mm. (b) Measured transmission magnitude $|S_{21}|$ and phase. (c) Measured plasma frequency $f_\mathrm{p}$ versus inverse finger length $1/L$. Symbols are measurements and solid lines are predictions from Eq.~(\ref{eq:fp}). (d) Simulated miniaturization factor as a function of cutoff frequency (the effective plasma frequency). }
\label{Fig3}
\end{figure}

The optical images in Fig.~\ref{Fig3}(a) show the comparison of two fabricated devices with the same cutoff frequency $f_\mathrm{p}\approx 5$~GHz: one on the K140 substrate ($\varepsilon=140$) and the other on the Rogers substrate ($\varepsilon=9.8$). Evidently the lateral size of the K140 device is roughly one-quarter of its Rogers counterpart. Both devices exhibit the characteristic SSPP response, as confirmed by the transmission spectra measured using a vector network analyzer (VNA) [Fig.~\ref{Fig3}(b). Black and gray: Rogers; Red and light red: K140]. For both devices, the magnitude $|S_{21}|$ shows a broad pass band terminated by a sharp cutoff at around 5 GHz, above which the transmission drops abruptly as the SSPP modes are no longer supported, while the phase accumulates rapidly near the cutoff, reflecting the reduced group velocity of the slow wave.

All six printed devices were characterized, exhibiting cutoff frequencies ranging from 2 to 11~GHz. Figure~\ref{Fig3}(c) shows the measured $f_\mathrm{p}$ as a function of $1/L$ (red circles). Evidently the $f_\mathrm{p}$ increases linearly with $1/L$, in great agreement with the theoretical prediction (solid red line) from Eq.~(\ref{eq:fp}). As a comparison, the same geometries are fabricated on the Rogers substrate, and the measured $f_\mathrm{p}$-$1/L$ relation is plotted in Fig.~\ref{Fig3}(c) (black squares), which also exhibits a perfect linear dependence and agrees with theoretical prediction (solid black line).
This linear scaling not only confirms the theoretical relation of Fig.~\ref{Fig1} in fabricated hardware but also provides a direct design chart from which the finger length needed for a target cutoff frequency can be read off for each substrate.

The experimentally determined finger lengths for the Rogers and K140 devices are shown as symbols in Fig.~\ref{Fig1}(d), validating the simulated permittivity-dependent scaling. From these measurements, a size-reduction factor of approximately four can be extracted for a given plasma frequency $f_\mathrm{p}$. According to Eq.~(\ref{eq:fp}), this factor is expected to scale as $\sqrt{\varepsilon_{\mathrm{eff,K140}}/\varepsilon_{\mathrm{eff,Rogers}}}$. Under the approximation $\varepsilon_{\mathrm{eff}} \approx \varepsilon$, substituting the permittivities of the K140 and Rogers substrates yields a predicted scaling factor of 3.8, which is slightly smaller than the experimentally observed value. This discrepancy is likely attributable to fabrication tolerances and other device imperfections. 

To examine whether this scaling law remains valid over a broader design space, we performed COMSOL simulations for SSPP waveguides with different finger lengths and plasma frequencies on both substrates. The results are summarized in Fig.~\ref{Fig3}(d). For designs with $f_\mathrm{p} > 2$~GHz, a nearly constant scaling factor of approximately 3.8 is obtained, in excellent agreement with the theoretical prediction. However, for $f_\mathrm{p} < 2$~GHz, the scaling factor gradually decreases, indicating that the miniaturization advantage afforded by the high-permittivity substrate becomes less pronounced. This behavior can be attributed to the weaker confinement of the SSPP mode at lower frequencies. As a larger fraction of the electromagnetic field extends into the surrounding air, the effective permittivity $\varepsilon_{\mathrm{eff}}$ deviates increasingly from the substrate permittivity $\varepsilon$, thereby reducing the validity of the simple scaling relation.


\begin{figure}[tb]
\includegraphics[width=\linewidth]{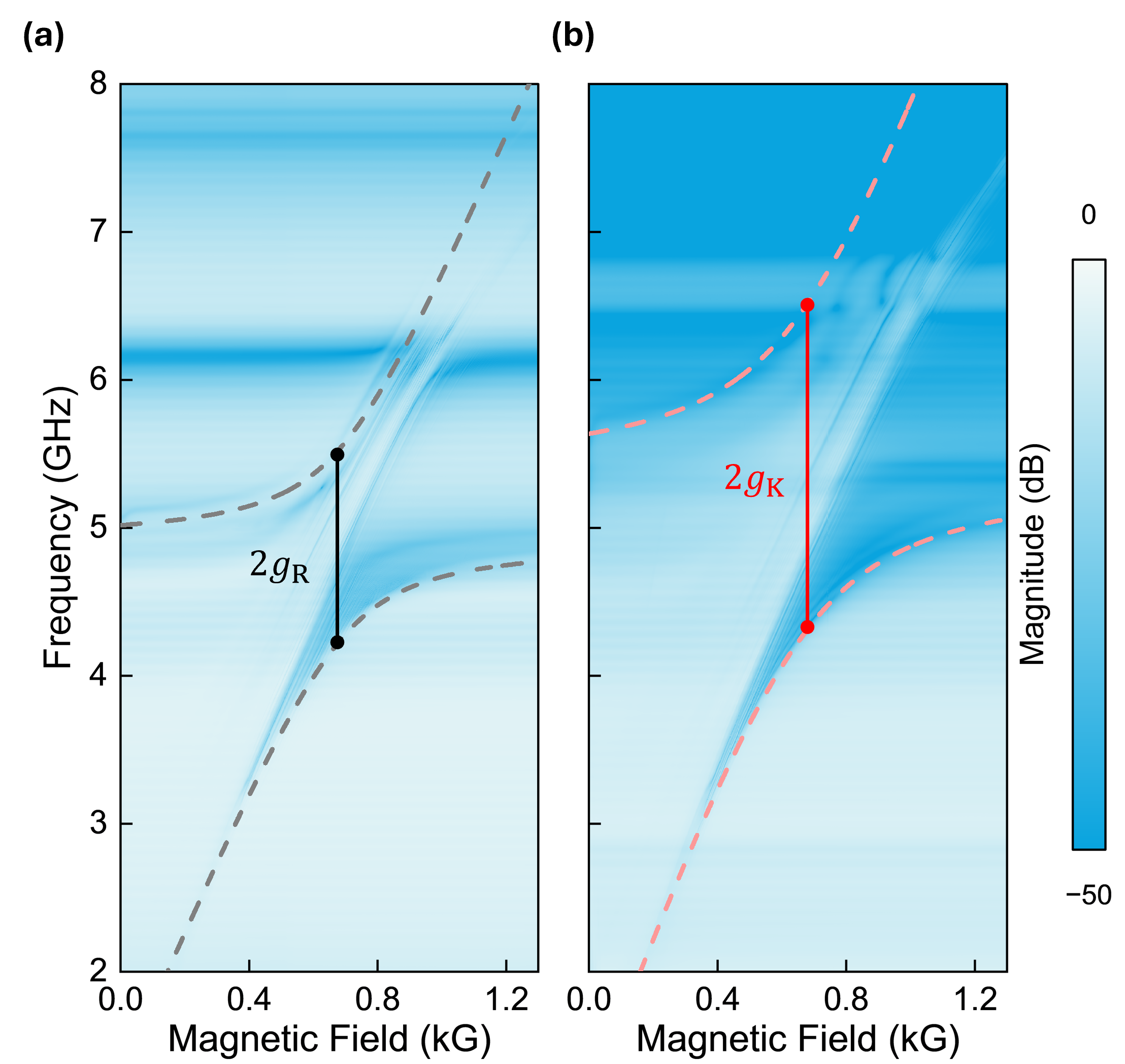}
\caption{ Magnon-SSPP coupling in Rogers and K140 devices. Measured transmission spectra as functions of frequency and bias magnetic field for SSPP waveguides fabricated on (a) Rogers ($\varepsilon=9.8$) and (b) K140 ($\varepsilon=140$) substrates. Both devices are loaded with a 23-$\mu$m-thick YIG film in the magnetostatic surface-wave configuration ($\mathbf{k}\perp\mathbf{H}$). Dashed curves are fits to the coupled-mode model, yielding coupling strengths of $g_{\mathrm{R}}=0.65$~GHz and $g_{\mathrm{K}}=1.1$~GHz. }
\label{Fig4}
\end{figure}

Beyond reducing the device size, the enhanced miniaturization provided by the high-permittivity substrate also leads to stronger electromagnetic-field confinement, which is expected to increase the magnon-photon coupling strength. To verify this effect, we compare the two SSPP devices fabricated on two different substrates but operating under identical magnonic conditions. Each waveguide is loaded with the same 23-$\mu$m-thick YIG film and biased by an in-plane magnetic field perpendicular to the SSPP propagation direction, corresponding to the magnetostatic surface-wave geometry ($\mathbf{k}\perp\mathbf{H}$). By sweeping the bias field, the magnon resonance is tuned across the SSPP band, producing two-dimensional transmission spectra as functions of frequency and magnetic field [Figs.~\ref{Fig4}(a) and \ref{Fig4}(b)]. In both devices, a pronounced avoided crossing is observed when the magnon resonance frequency is swept across the SSPP cutoff, providing clear evidence of coherent magnon-photon hybridization.

Fitting the hybridized modes to the coupled-mode model~\cite{2024_PRL_Xu_slowwave} yields coupling strengths of $g_{\mathrm{R}} = 0.65$~GHz for the Rogers device and $g_{\mathrm{K}} = 1.1$~GHz for the K140 device, as indicated by the dashed curves in Figs.~\ref{Fig4}(a) and \ref{Fig4}(b). Thus, under otherwise identical conditions, replacing the Rogers substrate with the K140 substrate increases the coupling strength by nearly a factor of two. This enhancement is consistent with the stronger electromagnetic-field confinement and reduced mode volume enabled by the high-permittivity SSPP waveguide.

This performance represents ultrastrong coupling and surpasses previous slow-wave magnon-photon coupling demonstrations. Reaching the ultrastrong-coupling regime previously required loading the SSPP waveguide with a bulk YIG disk (350~$\mu$m in thickness and 5~mm in diameter), whose large magnetic volume compensated for a relatively weak per-spin interaction to attain $g/f_\mathrm{p} = 19\%$~\cite{2024_PRL_Xu_slowwave}. In contrast, the device reported here employs a 23~$\mu$m-thick, $3\times3$~mm$^2$ YIG film whose magnetic volume is reduced by more than an order of magnitude, yet achieves $g/f_\mathrm{p} \approx 22\%$, pushing the system even deeper into the ultrastrong-coupling regime. This improvement indicates a substantially enhanced effective per-spin coupling, which originates from the reduced SSPP mode volume and the resulting concentration of the microwave field near the magnetic film. Ultrahigh-permittivity substrates therefore enable a markedly more compact slow-wave device while simultaneously pushing magnon-photon coupling further into the ultrastrong-coupling regime.


In summary, we have demonstrated a practical route toward compact slow-wave hybrid magnonic devices by combining ultrahigh-permittivity substrates with AME. A ceramic substrate with $\varepsilon=140$ reduces the SSPP finger length required for a given cutoff frequency by nearly a factor of four compared with conventional microwave substrates, consistent with theoretical predictions. Three-dimensional inkjet printing enables thick, low-loss conductors to be patterned directly onto brittle ceramic substrates through a maskless process that supports rapid design iteration. Beyond reducing the device footprint, the enhanced field confinement strengthens magnon–photon coupling, showing that miniaturization and strong coherent interactions can be advanced together in this platform. This work points to a unique direction for the development of integrated hybrid magnonic devices combining enhanced coupling, broad bandwidth, and compact footprints.

\begin{acknowledgments}
X.Z. acknowledges support from National Science Foundation (2337713). The authors gratefully acknowledge QTREX Quantum (Luis Unger and Jacky Borenstein) for their valuable technical guidance and support related to AME process parameters and design rules. 
\end{acknowledgments}


\bibliography{BBME}

\end{document}